\documentclass{emulateapj}
\usepackage{lscape}
\usepackage{natbib}

\shorttitle{Leo A}
\shortauthors{Cole et al.}

\begin{document}

\title{Leo A: A Late-Blooming Survivor of the Epoch of
Reionization in the Local Group}

\author{Andrew A. Cole\altaffilmark{1},
Evan D. Skillman\altaffilmark{1},
Eline Tolstoy\altaffilmark{2},
John S. Gallagher, III\altaffilmark{3},
Antonio Aparicio\altaffilmark{4,5},
Andrew E. Dolphin\altaffilmark{6},
Carme Gallart\altaffilmark{4},
Sebastian L. Hidalgo\altaffilmark{1}
Abhijit Saha\altaffilmark{7},
Peter B. Stetson\altaffilmark{8},
Daniel R. Weisz\altaffilmark{1},
}

\altaffiltext{1}{University of Minnesota, Department of Astronomy,
  116 Church Street S.E., Minneapolis, MN 55455, USA; cole, skillman,
  slhidalgo@astro.umn.edu, dweisz@astro.umn.edu.}
\altaffiltext{2}{Kapteyn Astronomical Institute, University of
  Groningen, Postbus 800, 9700~AV Groningen, Netherlands; etolstoy@astro.rug.nl.}
\altaffiltext{3}{Department of Astronomy, University of Wisconsin-Madison,
  5534 Sterling Hall, 475 North Charter Street, Madison, WI 53706, USA;
  jsg@astro.wisc.edu.}
\altaffiltext{4}{Instituto de Astrof\'{i}sica de Canarias, Calle
  V\'{i}a L\'{a}ctea, E-38200, La~Laguna, Tenerife, Spain; antapaj,
  carme@iac.es.}
\altaffiltext{5}{Department of Astrophysics, University of La~Laguna,
  Tenerife, Spain.}
\altaffiltext{6}{Steward Observatory, University of Arizona, 933 N.
  Cherry Ave., Tucson, AZ 85721, USA; Raytheon Corporation; adolphin@raytheon.com.}
\altaffiltext{7}{National Optical Astronomy Observatory, P.O. Box
  26732, Tucson, AZ 85726, USA; saha@noao.edu.}
\altaffiltext{8}{Dominion Astrophysical Observatory, Herzberg Institute
  of Astrophysics, National Research Council, 5071 West Saanich Road,
  Victoria, BC~V9E~2E7, Canada; peter.stetson@nrc-cnrc.gc.ca.}

\begin{abstract}
As part of a major program to use isolated Local Group dwarf galaxies
as near-field probes of cosmology, we 
have obtained deep images of the dwarf irregular galaxy
Leo~A with the Advanced Camera for Surveys aboard the Hubble Space 
Telescope.  From these images we have constructed a color-magnitude
diagram (CMD) reaching apparent [absolute] magnitudes of (M$_{475}$, M$_{814}$) 
$\gtrsim$ (29.0 [$+$4.4], 27.9 [$+$3.4]), the deepest ever
achieved for any irregular galaxy
beyond the Magellanic Clouds.  We derive the star-formation rate (SFR)
as a function of time over the entire history of the galaxy.
We find that over 90\% of all the star formation that ever occurred
in Leo~A happened more recently than 8~Gyr ago.  The CMD shows only
a very small amount of star formation in the first few billion years
after the Big Bang; a possible burst at the oldest ages cannot be
claimed with high confidence.  The peak SFR occurred
$\approx$1.5--4~Gyr ago, at a level 5--10 times the current value.
Our modelling indicates that Leo~A has experienced very little
metallicity evolution;
the mean inferred metallicity is
consistent with measurements of the present-day gas-phase oxygen
abundance.  We cannot exclude a scenario in which {\it all} of the
ancient star formation occurred prior to the end of the era
of reionization, but it seems unlikely that the lack of star
formation prior to $\approx$8~Gyr ago was due to early loss or exhaustion
of the {\it in situ} gas reservoir.
\end{abstract}
\keywords{galaxies: dwarf --- 
color-magnitude diagrams --- galaxies: evolution}

\section{Introduction}
\label{sec-int}

Dwarf galaxies are the most common class of galaxy in the
Universe, which gives them significance far beyond their mass.
They are simpler in structure and perhaps more easily understood
than giant galaxies, and so their study can shed light on the
processes governing galaxy evolution.  Dwarf galaxies are also
important to cosmology, because they are similar in mass to
predictions for the proto-galactic fragments that collapse
hierarchically to form giant galaxies.  One of the most
interesting testable predictions to come from numerical simulations
of galaxy formation in a cold dark matter (CDM) cosmology 
is that after the Universe is
reionized at redshifts $z$ $\gtrsim$6, galaxies of total mass
$\lesssim$10$^8$ M$_{\sun}$ are prevented from further accretion
of gas by photoionization feedback from the cosmic ultraviolet background
\citep[e.g.,][]{nav97}.
This {\it may} strongly impact a dwarf's evolution, but observational
evidence is inconclusive.

\begin{deluxetable}{lcl}
\tablecaption{Global Properties of Leo~A\label{tab-prop}}
\tablewidth{0pt}
\tablehead{
  \colhead{Quantity}&
  \colhead{Value}&
  \colhead{Reference}
}
\startdata
Galactic coordinates ($\ell$, $b$)  & 196.9, $+$52.4         &  1 \\
distance modulus (m$-$M)$_0$        & 24.5 $\pm$0.1          &  2 \\
reddening E(B$-$V)                  & 0.021                  &  3 \\
Absolute magnitude M$^0_B$          & $-$11.7 $\pm$0.2       &  4 \\
M(H{\tiny I}) (10$^7$ M$_{\sun}$)   & 1.1 $\pm$0.2           &  5  \\
Total Mass (10$^7$ M$_{\sun}$)      & $\lesssim$20           &  5  \\
12$+\log$(O/H)                      & 7.38 $\pm$0.1          &  6  \\
Holmberg semi-axes, $a_H$,$b_H$     & 3$\farcm$5, 2$\farcm$2 &  7  \\
Star-formation rate (M$_{\sun}$/yr) & 1--2$\times$10$^{-4}$  & 8,9 \\
\enddata
\tablerefs{(1) NED; (2) \citet{dol02b}; (3) \citet{sch98}; 
  (4) RC3; 
  (5) \citet{you96}; (6) \citet{van06}; (7) \citet{fis75};
  (8) \citet{hun04}; (9) \citet{jam04}.}
\end{deluxetable}

We are fortunate 
to have a plethora of dwarf galaxies available to
study in the Local Group; their proximity makes
it possible to measure their histories
in far more detail than is possible
in even the next-closest groups.  
Because the relative importance of mergers and interactions as compared
to intrinsic properties is not well-known, it is important to study dwarfs
in as many environments as possible.  Therefore we have begun a
Hubble Space Telescope
Advanced Camera for Surveys (ACS) program to measure the complete
star-formation
histories (SFH) of six isolated Local Group dwarf galaxies 
in order to search for the
imprints of cosmological processes on their evolution: the LCID 
(Local Cosmology from Isolated Dwarfs) project (Gallart et al.,
in preparation).  By measuring several small, isolated
galaxies, we will
test for correlations and patterns in their SFH that may provide clues
as to the magnitude of the effects of reionization, supernova blowout,
and other feedback processes.

Here we present our first results, a new measurement of the
SFH of the galaxy Leo~A (DDO~69).  
Leo~A was discovered by \citet{zwi42} in the course of 
a search for the lowest luminosity galaxies.  It is indeed one
of the least luminous gas-rich galaxies known \citep[e.g.,][]{mat98}.
The vital statistics of Leo~A are given in Table~\ref{tab-prop}.
Leo~A is a small, blue galaxy conspicuous for its population 
of young, massive stars, with estimated ages from
$\approx$10$^7$--10$^8$ yr \citep{dem84,tol96a}.
The star-formation rate (SFR) has apparently declined since
that time, although a few small H{\small~II} regions are present.

The first HST observations of Leo~A \citep{tol98}
revealed a very high fraction of stars aged $\approx$1--2~Gyr
at very low metallicity, suggesting that only $\sim$10\% of the
stellar mass was contained in older stars.
\citet{sch02} obtained slightly deeper data in an offset
field and measured a lower overall SFR and much higher fraaction
of ancient stars: they found evidence for a median age of $\approx$4~Gyr,
with the bulk of the old stars older than $\approx$10~Gyr.
Neither of these
studies reached the necessary depth 
to convincingly measure the fraction and
age distribution of the oldest stars.
\citet{dol02b} supplied the first proof of the existence of truly ancient
stars in Leo~A with the discovery of RR~Lyrae type variables.
Our goal here is to 
provide a definitive measurement of how Leo~A's SFR has varied
with time over its entire lifetime.
Further details, including spatial
patterns within Leo~A, comparison to numerical models, and thorough
comparison to other galaxies in our sample, will be deferred to
future papers, in preparation.

\section{Data Acquisition and Photometry}
\label{sec-obs}

\begin{figure}[t]
\epsscale{1.1}
\plotone{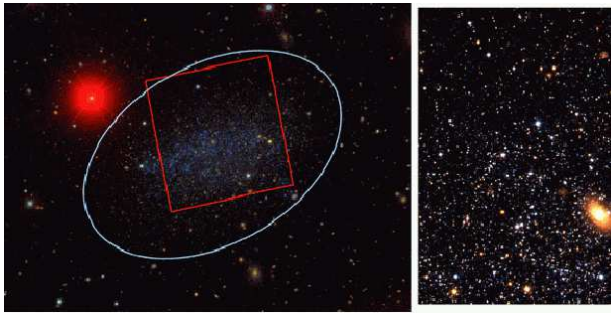}
\caption{{\it left:} Leo~A as seen in a mosaic of SDSS 
images (courtesy D.W. Hogg). The image is
10$\farcm$5 $\times$ 7$\farcm$8; North is up and East is to the left.
Our ACS/WFC field is marked in red and the Holmberg radius 
in cyan.
{\it right:} 36$\arcsec$ $\times$ 54$\arcsec$ detail of our ACS/WFC
field near image center.
Background galaxies are clearly visible through the body of
Leo~A.  The high ratio of bright blue to bright red stars is immediately
apparent, as is the north-south density gradient.
\label{fig-map}}
\end{figure}

Leo~A was observed with ACS
over 16 orbits during the time period
26 December 2005 to 8 January 2006.  Each HST orbit was 
devoted to a single exposure of at least 2400 seconds duration,
split to facilitate cosmic ray removal.  We chose the filter
pair F475W (Sloan $g^{\prime}$) and F814W (Broad I) as having
the best combination of throughput and temperature sensitivity
for F--G type dwarfs.  Our paramount concern in designing the
observations was to achieve the maximum possible depth of 
photometry, so that our conclusions would be as free as possible
from difficult-to-characterize systematics.  We used small dithers
between exposures to allow us to correct for hot pixels in the 
detector, and observed when the Sun-target angle was over 
100$^{\circ}$.  The total exposure time in (F475W, F814W) 
was (19,200, 19,520) seconds.  

The Leo~A field is shown 
in Figure \ref{fig-map}, where the left panel shows a mosaic
of images from the Sloan Digital Sky Survey
\citep[fourth data release;][]{ade06}, centered on the
galaxy.  The footprint of our ACS observation is marked by the
red square, while the cyan ellipse shows the approximate location
of the Holmberg radius.  It is important to remember
that a low surface density sheet of red giant stars extends out
to $\approx$7$\farcm5$ \citep{dol02b,van04}, 
and in the outer reaches of the galaxy the
ratio of young to old stars must be significantly lower than what
we find here for the center of Leo~A. 

The pipeline-processed data were combined using MultiDrizzle to 
eliminate cosmic rays, and photometered using DOLPHOT, a version
of HSTPHOT \citep{dol00} incorporating the best available ACS calibrations.
Extended objects and residual hot pixels were rejected based on their
brightness profiles, leaving 112,000 well-measured stars in the
color-magnitude diagram (CMD).  Artificial star tests
show that the typical photometric error reaches $\pm$0.1 mag for
(m$_{475}$, m$_{814}$) = (28.7, 27.9); the 50\% completeness limit
is at (m$_{475}$, m$_{814}$) = (29.0, 27.9).

\begin{figure}[t]
\epsscale{1.3}
\plotone{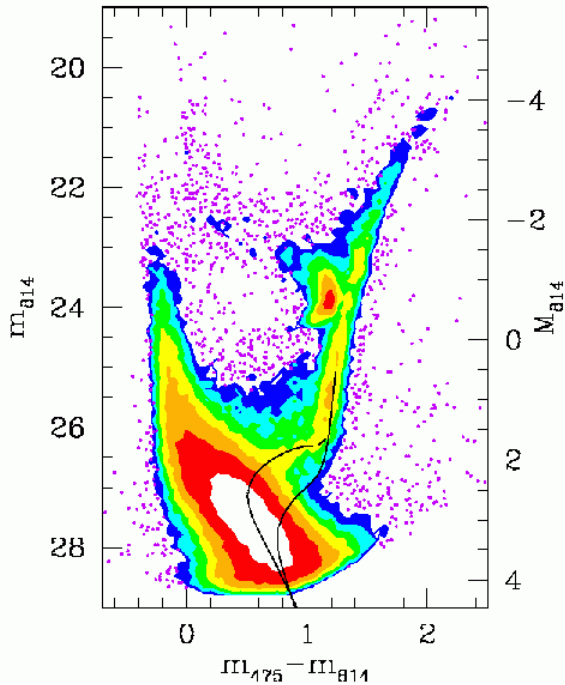}
\caption{Our ACS/WFC CMD for Leo~A.
Individual stars are plotted where their 
density is less than 8 stars/decimag$^{2}$.  Contours
are uniformly spaced by factors of two to show the 
overall density distribution and the fine structure of the 
stellar sequences. The black lines show the main-sequence and lower 
RGB of isochrones with (Z, age/Gyr) $=$ (0.0001, 14) and (0.001, 5),
respectively.
\label{fig-cmd}}
\end{figure}

The CMD is shown in Figure~\ref{fig-cmd}.  Where the density of stars
is high enough to confound the easy estimation of the relative number
of stars in the various sequences, we have plotted contours corresponding
to stellar density rather than individual stars.  The contours cover
the range from 8--512 stars/decimag$^{2}$, evenly
spaced by factors of 2.  This allows both the overall density distribution
as well as the fine structure of the principal sequences to be shown
in a single figure.  

The main features which bear on the measurement of
the star-formation history, especially at the previously ill-constrained
earliest times, are the relatively bright locations of the peak 
density of subgiants and the main-sequence turnoff; the underdeveloped
horizontal branch and strong, vertically extended red clump; and the
paucity of upper red giant branch stars.  The implication of the bright
subgiant branch is immediately evident by comparison to the location
of the old, metal-poor isochrone in Fig.~\ref{fig-cmd}: 14~Gyr old stars
would produce a far fainter subgiant branch than observed.  The peak
density of subgiants occurs at M$_{814}$ $\approx$ $+$1.9, more than
half a magnitude brighter than comparable measurements in globular clusters
of similar metallicity \citep[e.g.,][]{ros00}, but similar to intermediate-age
clusters in the Small Magellanic Cloud \citep[e.g.,][]{mig98} and well-matched
to isochrones with ages $\sim$5~Gyr.  
This is striking because a constant SFR produces a steep and monotonic
increase in subgiant density with age \citep[e.g., Fig.~1 of ][]{gal05}.
Because our 
photometry reaches the oldest main-sequence turnoff,
we can unambiguously quantify the galaxy's youth for the first time.

\section{Derivation of the Star-Formation History}
\label{sec-age}

The quantitative star-formation history of a stellar population can 
be derived from a CMD by comparison of the distribution of stars
predicted by an ensemble of theoretical models, corrected
for the effects of distance, interstellar reddening, and
the completeness and precision limits imposed by the 
observations \citep[e.g.,][]{tol96b}.  The model predictions 
are derived from theoretical isochrones plus a prescription
for the intital mass function and the
influence of unresolved binary stars.
The adopted isochrone sets have been created in the ACS/WFC filter system
(L. Girardi, private communication), so no transformation of our
photometry was required.  
The variation of SFR with time and the
evolution of stellar metallicity are taken from the combination
of parameters that results in the best reproduction of the
CMD.  The ``best-fit'' synthetic CMD is quantified by a
$\chi ^2$-like statistic derived from the definition of
the maximum likelihood for the case of Poisson-distributed
data \citep{dol02}.  

We adopt a simulated annealing approach to the minimization problem
\citep{kir83}, in order to find the best CMD without searching all 
of parameter space, while avoiding false local minima.  An example
of the type of local minima encountered in this problem is the
range of solutions produced by age-metallicity degeneracy.  Our
implementation of the procedure (by A.A.C.) is described in 
\citet{ski03}.  Our recent tests using various independent 
algorithms have proven that for data of high quality, the details
of the fitting procedure do not strongly affect the derived
SFH
\citep[see also][and Aparicio \& Hidalgo, in preparation]{ski02,ski03}.

The critical parameters (m$-$M)$_0$ and E(B$-$V) were taken from 
the sources in Table~\ref{tab-prop}.  The solutions shown here were
obtained using isochrones from \citet{gir00} and the IMF from 
\citet{cha03}.
Extinction in the ACS/WFC filter system was obtained by interpolation
within the tables of \citet{sir05}: A$_{814}$ = 0.039, A$_{475}$ = 0.078.

We solved for the SFH in 9 age bins; the age bins are logarithmically
spaced to reflect the physical reality that the separation between
isochrones of different ages is a strongly decreasing function of age.
Within each age bin, a range of metallicities was allowed, from the
lowest available metallicity to a value slightly higher than that 
corresponding to the \citet{van06} metallicity: 0.0001 $\leq$ Z
$\leq$ 0.0015.  The only constraint on age-metallicity relation 
was to exclude the lowest metallicity isochrones from the youngest
age bins, as they are seen by inspection to be mismatched to the data.
To restrict ourselves to inferences drawn from the best-modelled
phases of stellar evolution, we use only the main-sequence
and subgiant branch in our likelihood calculation. This does not
significantly change the derived SFH because we include $>$90\%
of all stars, most of the excluded stars have very little 
age-sensitivity, and their colors are strongly model-dependent.

\begin{figure}[t]
\epsscale{1.1}
\plotone{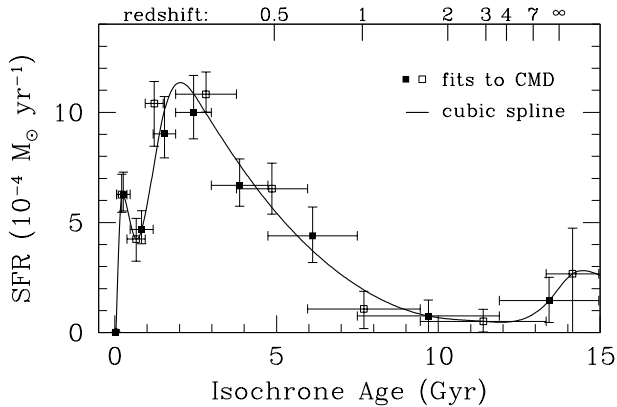}
\caption{The derived SFH of central Leo~A.
Data points with 1$\sigma$ errorbars show
the best fits to the CMD for two
age binning schemes (solid squares, open squares).  
The heavy line shows the results of a spline fit to
the results, our best estimate of the form of Leo~A's SFH.
If isochrone age $\equiv$ lookback time,
the redshift derived using the third-year WMAP results
and flat $\Lambda$CDM cosmology is given along the top axis 
(note that the isochrone age scale and the WMAP age of the Universe
are incongruent; see text for details).
\label{fig-sfh}}
\end{figure}

Our derived history of star formation in Leo~A is shown in 
Figure~\ref{fig-sfh}.  The best-fit SFH is shown for two 
different age binnings-- solid squares and open squares, respectively.
Each point carries its 1$\sigma$
errorbar in SFR and an age errorbar marking the width of the age bin.
The heavy curve is a cubic spline fit to the results; it represents
our best estimate of the form of the 
SFH in the central part of Leo~A.
Because the total number of stars is fixed by the data, increasing the
SFR in one age bin demands a decrease 
in the adjacent bin(s).  Consequently, the
errorbars in Fig.~\ref{fig-sfh} imply that a slightly older, broader
SFR peak is allowed by the data.
The derived rates apply to the ACS/WFC field of view,
which covers 0.6~kpc$^2$ at a distance of 795~kpc; this is
$\approx$47\% of the area within the Holmberg radius, and 
$\approx$29\% of the area of Leo~A's intermediate-age disk
\citep{van04}.  The redshift scale at top applies to a flat
$\Lambda$CDM
universe with parameters taken from the WMAP year-3 dataset 
\citep{spe06}.  The stellar evolution and cosmological clocks
disagree at early times because of missing physics in the 
stellar models.  The most important of these is gravitational
diffusion settling of light elements, which reduces the ages
of low-mass stars by 10--20\% \citep{pro91,cha92}.

Both solutions were obtained with fixed values of distance modulus,
reddening, and binary star properties.  We tested the sensitivity
of the derived solution to reasonable variations in these parameters,
and found that
for all reasonable parameter choices, the derived
SFH has the same general form, because of the exceptional
photometric depth and completeness (Cole et al., in preparation). 
Leo~A is best described as a late-blooming galaxy, in
which the great majority of stars are formed after a delay of
several Gyr.  The apparent early peak of star formation above a
subsequent minimum is not a high-confidence feature, as the 
errorbars indicate.

The derived metallicity evolution is not plotted in Figure~\ref{fig-sfh}
because there is almost no trend with age.  A constant heavy element mass
fraction Z = 0.0008$^{+0.0005}_{-0.0003}$ ([Fe/H] $\approx$ $-$1.4 $\pm$0.2;
ranges represent the rms scatter at a given age)
is an adequate representation of the results, although there is some hint
of a trend of increasing metallicity with time, from Z = 0.0006 to Z = 0.001.

\section{Summary \& Discussion}
\label{sec-sum}

As suggested by \citet{tol98}, Leo~A is a predominantly
young galaxy.  However, the depth of our photometry allows
us to much more precisely describe its stellar age distribution:
within the ACS/WFC field of view, 90\% of the star formation 
has occurred more recently than 8~Gyr ago.  The
current star formation rate of $\approx$10$^{-4}$ M$_{\sun}$
yr$^{-1}$ is in good agreement with the rates estimated
from H$\alpha$ emission.  The large number of stars above
the 5~Gyr isochrone shows that the SFR increased dramatically
around this time and continued at high level for $\approx$3~Gyr
before declining.
However, {\it some} truly ancient
stars are required to reproduce the observed CMD and 
provide the observed RR~Lyraes.  The 14~Gyr isochrone
in Fig.\ \ref{fig-cmd} traces the lower envelope of subgiant
stars, confirming that not {\it all} of the star formation
was delayed.

Integrating over the radial profile from \citet{van04}, 
we find that Leo~A only astrated
$\sim$1--2$\times$10$^{6}$ M$_{\sun}$ at ages corresponding
to redshifts $z > 1$. 
If all of its present-day baryon
inventory was present at high redshift, then Leo~A would
have been 95\% gas and only 5\% stars prior to $z\approx$2.
Although extreme, the M(\ion{H}{1})/L$_B$ ratio of the young
Leo~A would have been comparable to those of the most gas-rich
dwarf irregular systems seen today \citep[e.g.,][]{war04}.
Evidently some small galaxies can retain or build large gas
reservoirs without much depletion by star formation.
This makes Leo~A-like objects 
intriguing candidates for proto-galactic fragments that 
merge into giant galaxies, prompting bursts of star formation
without leaving much stellar residue in their halos.

Because of uncertainties in assigning {\it absolute} ages
to old stellar populations, and because our time resolution
decreases with age, we cannot rule out that all of the ancient
star formation in Leo~A took place prior to reionization.
That would require compressing all of the ancient star formation in 
Fig.~\ref{fig-sfh}, plus the formation of the unobserved Pop~III
stars that enriched the galaxy to its ``initial'' metallicity
[Fe/H] $\approx$ $-$1.5, into an $\sim$0.5~Gyr window.

Leo~A is not the only ``late bloomer'' in the Local Group, but
it appears to be an extreme case.  An example
of a qualitatively similar SFH can be found in the 
Leo~I \citep{gal99} dwarf spheroidal (dSph) galaxy.  The
two galaxies are of comparable mass, but the latter is a gas-free,
bound satellite of the Milky Way, with plausible triggers for
late star formation in tidal interactions or accretion events.
A trigger in the case of Leo~A is harder
to identify.

The idea that a dwarf galaxy's SFR is tied only
to the gas infall rate would require an 
unusual merger history for this tiny, isolated system.  
Leo~A would have gained virtually all of
its gas in the past few Gyr: not enough time for it to have
gotten to its present remote location from the vicinity
of M31 or the Milky Way where encounters might be expected to 
be more likely.  Given its probable long history of isolation,
it may be that much of the gas in Leo~A was present from
early on, and therefore the suppression of early star formation
is not attributable to the loss or exhaustion of its initial reservoir.
It seems more plausible
that only a small fraction of the \ion{H}{1} was able
to participate in star formation, with the rest kept warm
in a halo; the UV background or supernova feedback
are candidate heat sources.  Sequestering most of the gas
in a halo also speeds the chemical evolution clock,
which is interesting given the prompt enrichment and 
subsequent flat age-metallicity relation in the galaxy.
The warm halo would have been diffuse and metal-poor,
resulting in a long cooling timescale and a possible delay
before it could participate in star formation--
perhaps triggered by
a rare infall or interaction event in this isolated dwarf.

\acknowledgments

Support for this work was provided by NASA through grant GO-10590 from
the Space Telescope Science Institute, which is operated by AURA, Inc.,
under NASA contract NAS5-26555.  Thanks to David Hogg for creating the
SDSS mosaic image used in Figure~\ref{fig-map}.
AAC would like to thank Henry Lee, Mario Mateo, and 
Lucio Mayer for helpful discussions.

{}

\end{document}